\documentclass[a4paper]{article}
\usepackage{INTERSPEECH2022}
\usepackage{algorithm}
\usepackage{algorithm}
\usepackage{algorithmicx}
\usepackage{algpseudocode}
\usepackage{amsmath}
\usepackage{multirow}
\title{Adversarial Attacks on Machinery Fault Diagnosis}
\name{Jiahao Chen, Diqun Yan}
\address{
  Ningbo University, China}
\email{196003641@nbu.edu.cn, yandiqun@nbu.edu.cn}

\begin{document}

\maketitle
\begin{abstract}
  Despite the great progress of neural network-based (NN-based) machinery fault diagnosis methods, their robustness has been largely neglected, as they are easily fooled by adding imperceptible perturbation to the input. For fault diagnosis problems, we reformulate various adversarial attacks and investigate them under untargeted and targeted conditions. Experimental results on six typical NN-based models show that their accuracies are greatly reduced by adding small perturbations. We further propose a simple, efficient and universal scheme to protect victim models using data normalization. This work provides an in-depth look at adversarial examples of machinery vibration signals so as to develop protection methods against adversarial attacks and improve the robustness of NN-based models.
\end{abstract}
\noindent\textbf{Index Terms}: fault diagnosis, adversarial attack, data normalization

\section{Introduction}

\begin{figure}[t]
  \centering
  \includegraphics[width=\linewidth]{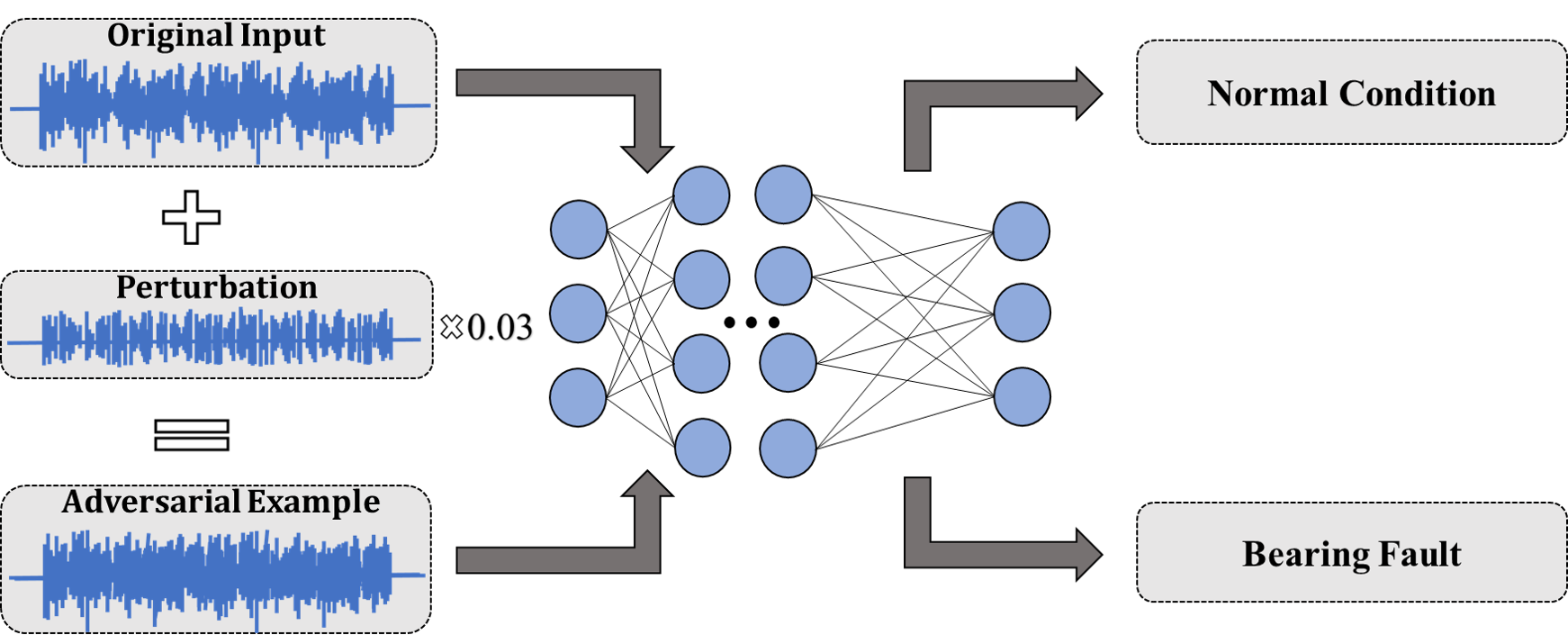}
  \caption{Schematic diagram of speech production.}
  \label{fig:pic1}
\end{figure}

Fault diagnosis has been applied in autopilots, aero engines, and wind energy conversion systems, and aims to diagnose the faulty part of the machinery equipment by finding abnormal vibration signals \cite{ref1,ref2,ref3}. Machinery fault diagnosis models can generally be classified as model-based, signal-based, or knowledge-based, or as hybrid/active methods \cite{ref4}. Knowledge-based methods such as deep neural networks (DNNs), have been widely investigated for their ability to establish explicit models or signal symptoms for complex systems \cite{ref5}, and have gradually replaced traditional knowledge-based methods such as support vector machines, etc. However, DNNs are vulnerable to adversarial attacks \cite{ref6}. Therefore, investigating the robustness of fault-diagnosis models is important, for the following reasons: (i) As can be seen from Figure \ref{fig:pic1}, even if the additional perturbations are unperceivable, the specific model can be easily cheated. (ii) Adversarial examples of obscure vibration signals must be investigated to find countermeasures to detect intentional attacks. The former reason helps defend against the attack through adversarial training, while the latter reason can help to identify the adversarial examples to guarantee the performance of the machinery equipment in various fields \cite{ref7,ref8,ref9}.

The concept of adversarial examples was proposed in the image domain and expanded to audio \cite{ref10,ref11,ref12}. In the above scenarios, attacks must be camouflaged as normal examples by minimizing the perturbation, for it is possible that additional perturbation is perceived by the human eye or ear. However, the vibration signals differ, because the crucial information within them is hard to extract via the human ear. While there are many studies on fault diagnosis with the vibration signals of motors in vehicles or industrial equipment, adversarial attacks on vibration signals have not been studied.

The results of this paper show that adversarial attacks can easily cause the models to misclassify the input signals with limited signals, which indicates that the credibility of these models is not guaranteed in the real scene. Considering this, applications of fault diagnosis such as in chemical processes, power networks, electric machines, and industrial electronic equipment are the most common targets of the malicious attacks, but the defensive method to tackle with these attacks is hardly investigated before.

We propose an efficient and universal countermeasure for this situation according to the results of our experiments. For fault-diagnosis systems, if they can defend against human deliberate attacks, then their defense ability to random perturbation from environment will also be greatly improved. Additionally, the generated adversarial examples can also play a role in enhancing and evaluating the robustness of fault-diagnosis systems through adversarial training. We investigate the features of vibration signals under the perspective of adversarial attacks. The features of vibration signal are different from that of the voice signals, which have been intensively studied.

Vibration signals differ from voice signals, in the following aspects: (i) A sequence of machinery vibration signals embodies the features including frequency, periodicity, kurtosis factors, and crest factor, which are meaningful in fault-diagnosis, but lack the complex information such as emotional factor, voiceprint, and language difference \cite{ref13}. (ii)The close connections between the sampling points of the voice can gradually restrict performance of the attack models. But when it comes to the signals made by the motors, the result shows that there may be a growth spurt of the signals and many features vary with the change of the physical properties and the severity of the fault \cite{ref14}.

The contributions of this work are summarized as follows: (i) We reformulate the adversarial examples of vibration signals which has hardly been mentioned before. (ii) We redefine the distortion measure of such signals which are obscure to human beings. (iii) We intensively investigate various adversarial attacks under untargeted and targeted conditions. (iv) We further analyze the experimental results and propose a simple, and efficient method to protect victim models.

\section{Related Works}

In this section we describe the popular dataset and models of fault diagnosis. We first describe the features and types of the used dataset. And then we move on to discuss different fault diagnosis models that perform well in fault diagnosing.

\subsection{Fault Diagnosis Datasets}
There are many fault diagnosis datasets. The Bearing dataset, from Case Western Reserve University (CWRU) bearing data center, is completed by Case Western Reserve University. As the most widely used standard dataset for bearing vibration signal processing and fault diagnosis, the fault features of CWRU Bearing Datasets are obvious and the related references are abundant. In this work, we used the Drive End (DE) part of the CWRU dataset, with a 12KHz sampling rate, whose 10 categories include 9 faulty types and 1 normal type. The 9 faulty types, as shown in Table \ref{tab:tab1}, indicate the fault diameter, from 0.007 to 0.021 inches, and the faulty location of a bearing, inner race, ball, or outer race.

\begin{table}[t]
  \caption{Fault features to diagnose.}
  \label{tab:tab1}
  \centering
  \begin{tabular}{cccc}
    \toprule
    \textbf{Diameter} & \textbf{Inner Race} & \textbf{Ball} & \textbf{Outer Race} \\
    \midrule
    \textbf{0.007} & IR007 & B007 & OR007 \\
    \textbf{0.014} & IR014 & B014 & OR014 \\
    \textbf{0.021} & IR021 & B021 & OR021 \\
    \bottomrule
  \end{tabular}
\end{table}

\subsection{Fault Diagnosis Models}
\begin{figure}[t]
  \centering
  \includegraphics[width=\linewidth]{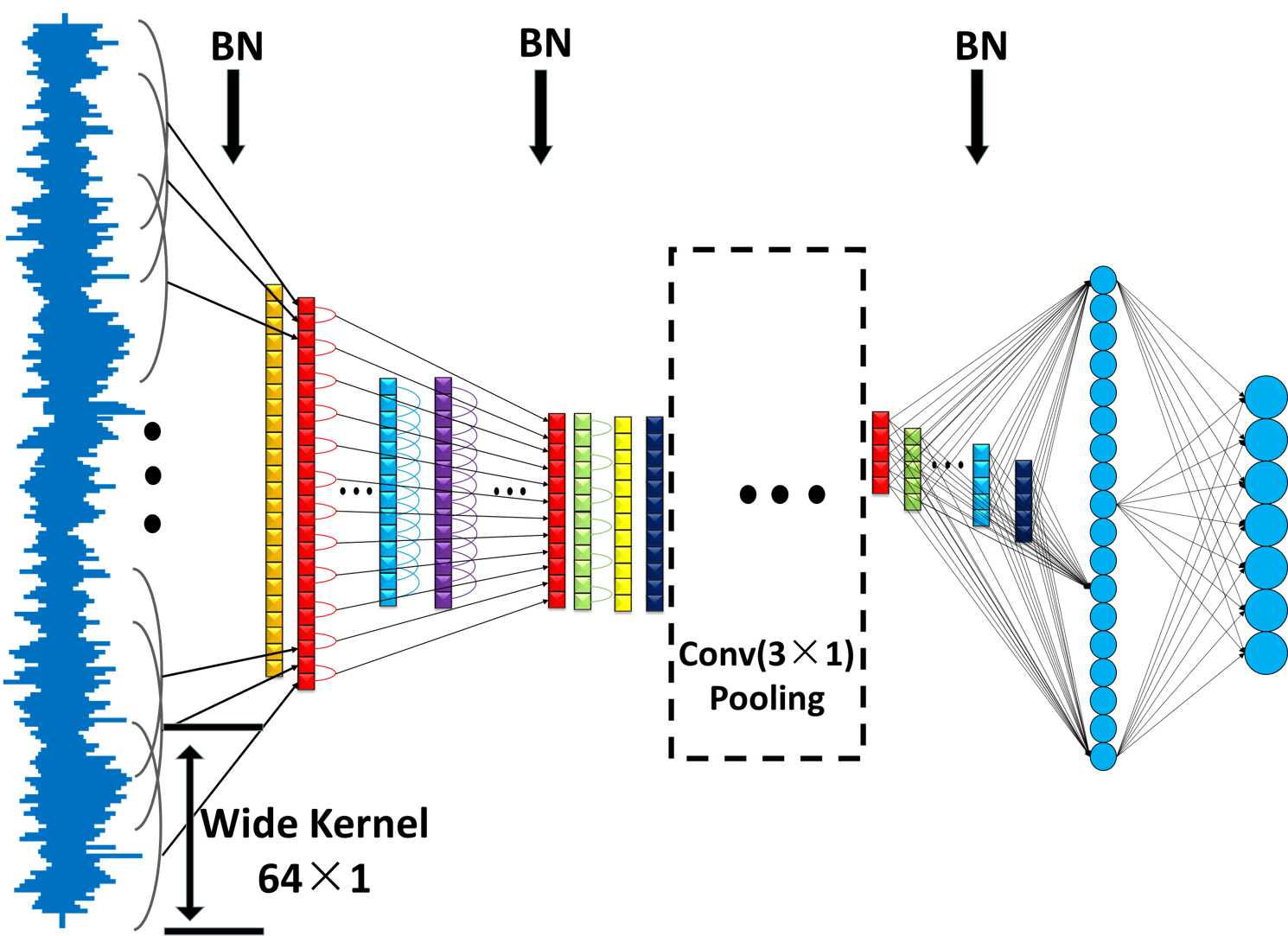}
  \caption{Structure of WDCNN.}
  \label{fig:pic2}
\end{figure}

Zhang \cite{ref15} used a convolutional neural network named WDCNN shown in Figure \ref{fig:pic2}, in which the first layer convolutional kernels were large, and the rest are small, to diagnose faults on CWRU Bearing dataset. This pattern allows the model to focus on the global features and reduces training time by avoiding a large number of convolutional layers. Additionally, batch normalization (BN) makes the model easy to train.

Turker Ince \cite{ref5} applied an improved CNN to the raw data (signal), eliminating the need for a separate feature extraction algorithm. This method is efficient in terms of speed and hardware. The experimental results demonstrated its effectiveness in monitoring motor conditions in real time, compared with traditional machine learning methods.

M. Zhao \cite{ref16} developed a variant of deep residual networks (DRNs), the so-called deep residual networks with dynamically weighted wavelet coefficients to improve diagnostic performance, which takes a series of sets of wavelet packet coefficients on various frequency bands as an input.

Some other models, also work well on CWRU Bearing dataset. For example, LeNet, AlexNet, BiLSTM can also achieve the accuracy of more than 99\% \cite{ref17}. All the above models mentioned acted as victim models in our experiment to evaluate the robustness of fault diagnosis.

\section{Methodology}
\subsection{Patterns of Adversarial Example}
Existing methods can be categorized as, gradient sign-based, optimization-based, evolutionary-based, or generate adversarial networks (GAN)-based \cite{ref18,ref19,ref20}. We used gradient sign-based methods to generate adversarial examples. Also, with different assumptions, background information and restrictions, we can divide the methods based on the following three aspects.

\textbf{White-Box and Black-Box.} For white-box, information on models, including the dataset, network architecture, weights and hyperparameters of the models, is accessible. While in most scenarios, it’s impractical. And for black box case, only the output of the model can be exploited. In this paper, we assume that the information of the models is accessible.

\textbf{Untargeted and Targeted.} Untargeted attacks aim to make the attacked models misclassify the input. And targeted attacks have a specific requirement for the output, which means that with additional perturbations, the models must transcribe the input to the targeted class. In fault-diagnosis conditions, the aim is to change the condition of the output from normal to faulty or vice versa. In this paper, both patterns were considered.

\textbf{Universal and Individual.} A perturbation generated for all samples functions universally for the whole dataset. Most of the existing attacks focus on individual attack based on a specific input. In this paper, the individual perturbation is used to fool the fault diagnosing classifiers.

\subsection{Adversarial Attack against Fault Diagnosis}
\textbf{Fast Gradient Sign Method (FGSM).} Goodfellow et al. proposed the approach, which can generate untargeted adversarial examples \cite{ref21}. Given the vibration signals $x$ of the fault-diagnosis system and its corresponding faulty type $y$, perturbation $\delta$ can be expressed as

\begin{equation}
  \delta=\operatorname{\varepsilon sign}\left(\nabla_{x} J(\theta, x, y)\right)
  \label{eq1}
\end{equation}
where $\theta$ denotes the parameters of fault-diagnosis system, and $J(\cdot)$ is the loss function. Given the target faulty type $y^{'}$, this method can also generate targeted adversarial examples by adding a perturbation

\begin{equation}
  \delta=-\varepsilon \operatorname{sign}\left(\nabla_{x} J\left(\theta, x, y^{\prime}\right)\right)
  \label{eq2}
\end{equation}

\textbf{Projected Gradient Descent (PGD)}. FGSM iterates for once, while PGD adopts many small iterations

\begin{equation}
  \begin{gathered}
x_{0}=x \\
x_{t+1}=\operatorname{clip}\left(x_{t}+\alpha \operatorname{sign}\left(\nabla_{x} J\left(\theta, x_{t}, y\right)\right)\right.
\end{gathered}
  \label{eq3}
\end{equation}
where $clip(\cdot)$ means that the perturbation must be restrained within the required scope from $1$ to $0$. Also, PGD can similarly obtain targeted faulty type \cite{ref22}.

\section{Experiment and Evaluation}
In this section we first introduce the experimental setup including the redefine of distortion measure and data processing. In our experiments we assume that we are accessible to the models and we conducted untargeted and targeted attack. Finally, we proposed our defensive method according to the experimental results.
\subsection{Experimental Setup}
\textbf{Distortion Measure.} For image and audio, whose information can be directly understood by humans, the distortion measures include the signal-to-noise ratio (SNR) and $L_2$ distance can guarantee the perturbations as imperceptible as possible. Since vibration signals are intricate, traditional distortion measures for adversarial examples are not applicable. The way to add noise restricts the operation of the attacks in fault-diagnosis condition is how to add noises. In reality, noises may emerge in the following forms: (i) The mechanical movement of electrical appliances. (ii) The circuit of the equipment within the system. (iii) The malicious attack from the computer virus to change the original data. (iv) The physical attack by imposing an external force on the sensors. Therefore, the attack cost can be understood as the external energy exerted on the vibration source and thus the energy $E(s)$ of the signals can be expressed as

\begin{equation}
  E(s)=\sum|x(k)|^{2}
  \label{eq4}
\end{equation}
where $x(k)$  denotes the sampling points of the origin signals. We define the measure of the attack cost as

\begin{equation}
  \operatorname{Cost}\left(x^{\prime}\right)=\operatorname{mean}\left(\log _{10}\left(\frac{\sum_{k=s}^{S+s-1}|x(k)|^{2}}{\sum_{k=s}^{S+s-1}|n(k)|^{2}}\right)\right)
  \label{eq5}
\end{equation}
where $x^{'}$ is the generated examples, $n(k)$ is the sequence of the noise, $S$ is the size of the segment, and $s$ is the start location of each segment. This measure denotes the degree of the attacks with maliciously designed or just environmental perturbations.

\begin{figure}[t]
  \centering
  \includegraphics[width=\linewidth]{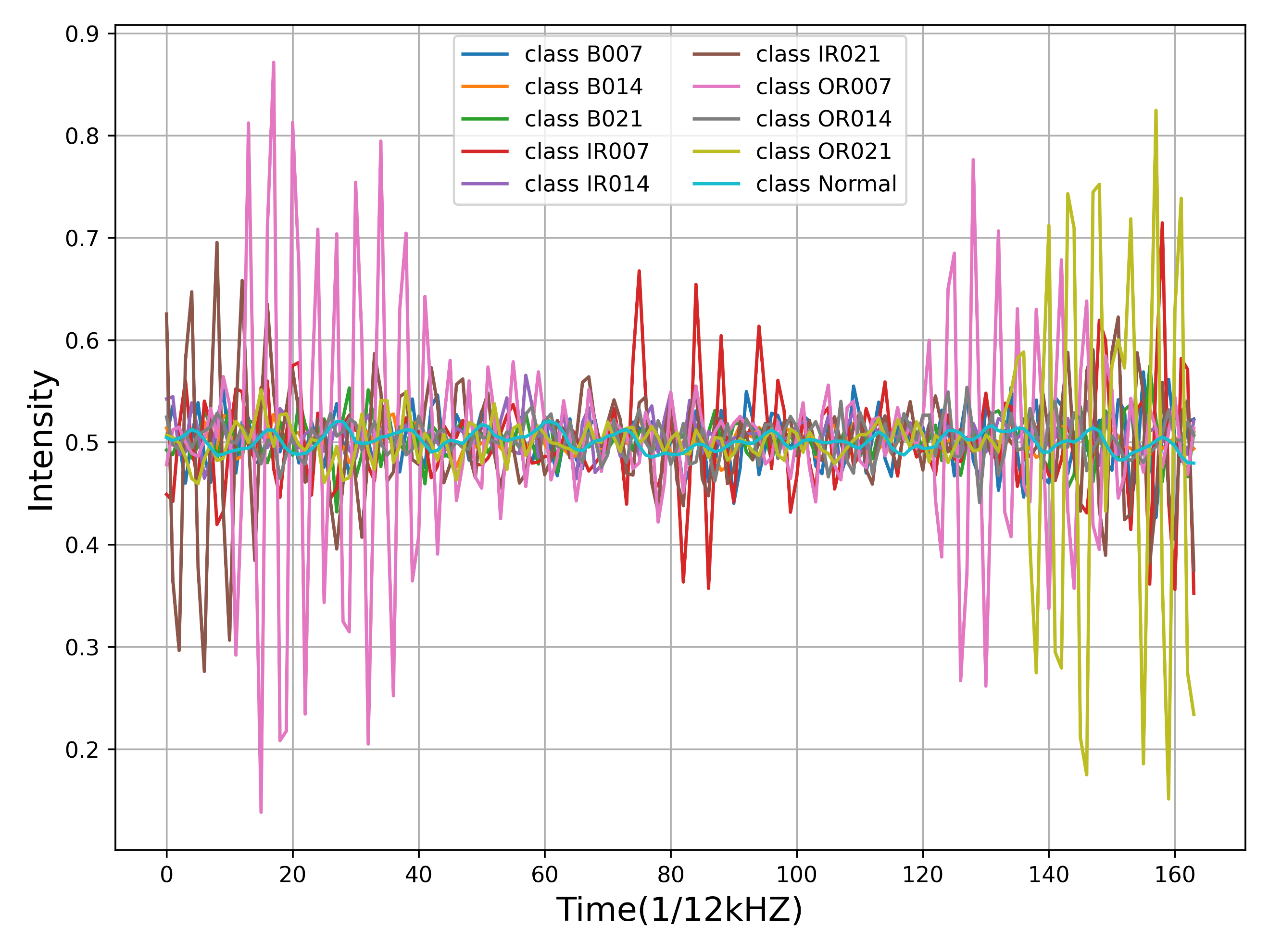}
  \caption{Vibration signals after data processing.}
  \label{fig:pic3}
\end{figure}

\textbf{Data Process.} The dataset was divided into training, validating and testing part with the ratio of $0.6, 0.2, 0.2$. In our experiment, each class has 1000 examples and each example has $2048$ normalized sampling points to make the distributions of all data similar and models converge quickly. In this paper, we improved the data processing used by Zhang to constrain the values of the signals between 0 and 1 \cite{ref15}. The data after processing is shown in Figure \ref{fig:pic3}, where there are some distinguishable differences between some classes, for example, class Normal and class $OR007$. Therefore, it is reasonable that the attack success rate from the faulty to normal is low, while the conversion between different types of faulty signals can be easily accomplished with a high success rate.

\subsection{Adversarial Attack on Fault Diagnosis}

\textbf{Pretrained Victim Models.} First, we pretrained the victim models to get the basic accuracy. After 1000 epochs’ training, all of the models can achieve the accuracy of more than $99\%$ on testing dataset, where the worst is $99.43\%$. Therefore, it’s obvious that the models can fit the dataset well.

\begin{figure}[t]
  \centering
  \includegraphics[width=\linewidth]{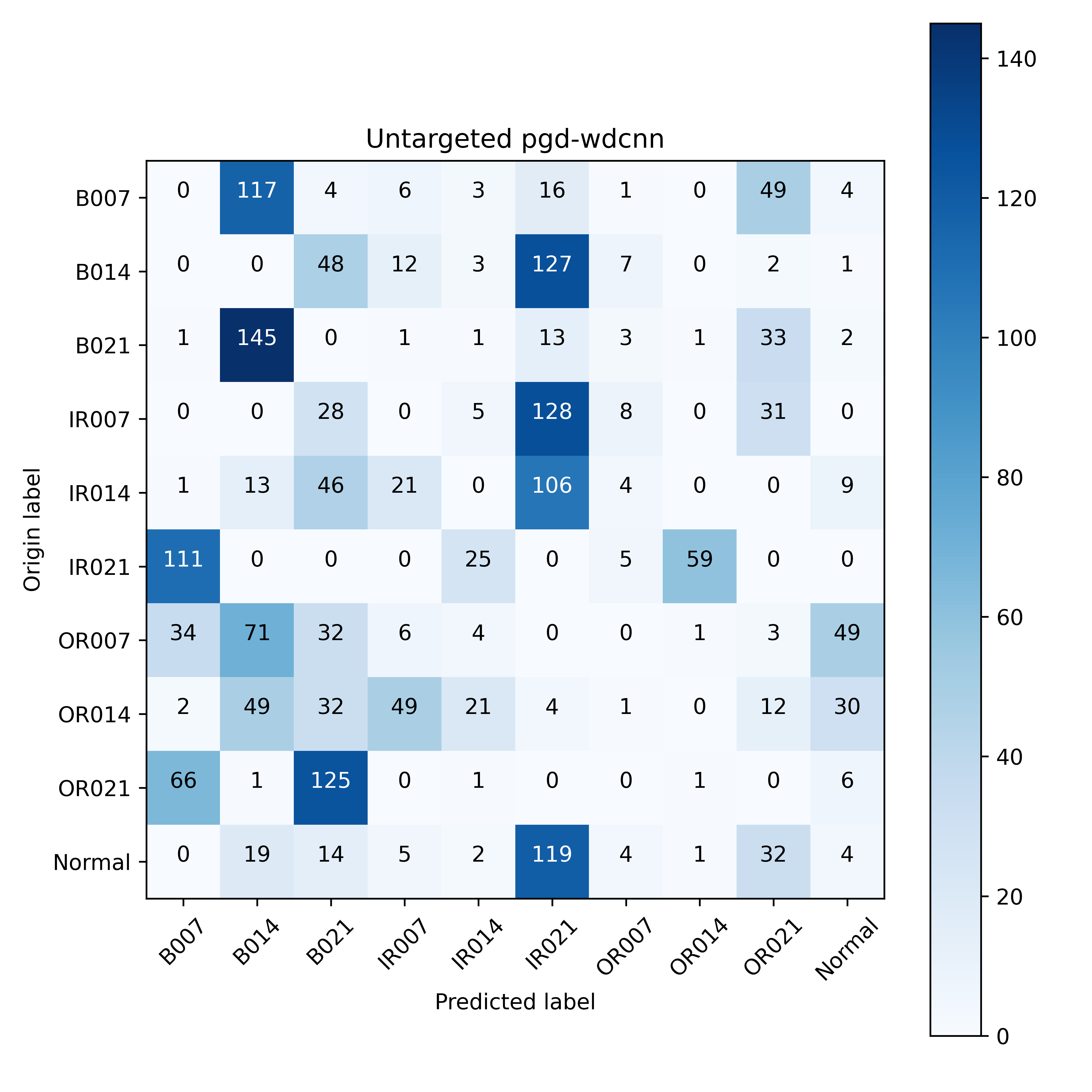}
  \caption{Confusion matrix of untargeted attacks.}
  \label{fig:pic4}
\end{figure}

\textbf{Untargeted Attack.} We generate untargeted adversarial examples with the test dataset to misclassify the models. The success rates of FGSM, and PGD are shown in Table \ref{tab:tab2}. As we can see in Table \cite{tab:tab2}, models such as WDCNN can be attacked with high accuracy and restricted cost. Hence, it is justifiable to think that DNNs-based fault diagnosis models are vulnerable to adversarial attacks. Additionally, from the confusion matrix given in Figure \ref{fig:pic4}. Instead of having random distribution, the misclassified classes were partial to certain classes such as $B014, B021, IR021$ and $OR021$. The partial predicted classes are marked with dark blue, which shows that the conversion difficulty between different types varies largely. 

\begin{table}[t]
\caption{Success rate of untargeted attacks.}
  \label{tab:tab2}
  \centering
\begin{tabular}{ccccccc}
\toprule
\multirow{2}{*}{\textbf{Models}} & 
\multicolumn{3}{c}{\textbf{FGSM}} & 
\multicolumn{3}{c}{\textbf{PGD}} \\
& \textbf{Mean} & \textbf{Best} & 
\textbf{Cost} & \textbf{Mean} & 
\textbf{Best} & \textbf{Cost}  \\
\midrule
\textbf{WDCNN} & 97.50 & 100 & 
1.72 & 99.90 & 100 & 0.73  \\
\textbf{LeNet} & 79.95 & 93.75 & 
1.72 & 99.95  & 100 & 0.66  \\
\textbf{DRNs} & 92.25 & 98.44 & 
1.72 & 95.20 & 100 & 0.61  \\
\textbf{AlexNet} & 65.80 & 81.25 & 
1.55 & 96.40 & 100 & 0.83  \\
\textbf{CNN1d} & 85.55 & 92.19 & 
1.72 & 94.25 & 98.40 & 0.69  \\
\textbf{BiLSTM} & 81.45 & 89.06 & 
1.93 & 92.15 & 100 & 0.88 \\
\bottomrule
\end{tabular}
\end{table}

\textbf{Targeted Attack.} We generated a perturbation to fool the models with targeted attack methods, with the results shown in Table \ref{tab:tab3}. The attack success rate varied greatly across models, but the success rate of one model, AlexNet, was extraordinarily higher than others. Based on this discovery, we proposed a simple scheme to defend against the malicious attacks.

\begin{table}[t]
\caption{Success rate of targeted attacks.}
  \label{tab:tab3}
  \centering
\begin{tabular}{ccccccc}
\toprule
\multirow{2}{*}{\textbf{Models}} & \multicolumn{3}{c}{\textbf{FGSM}} & \multicolumn{3}{c}{\textbf{PGD}} \\
 & \textbf{Mean} & \textbf{Best} & \textbf{Cost} & \textbf{Mean} & \textbf{Best} & \textbf{Cost}  \\
 \midrule
\textbf{WDCNN} & 15.85 & 21.88 & 1.72 & 29.55 & 37.50 & 1.03  \\
\textbf{LeNet}  & 0.05 & 1.56 & 1.72 & 9.95 & 20.30 & 1.12  \\
\textbf{DRNs} & 20.30 & 28.125 & 1.72 & 31.25 & 37.50 & 1.38  \\
\textbf{AlexNet} & \textbf{12.75} & \textbf{21.88} & \textbf{1.73} & \textbf{96.50} & \textbf{100} & \textbf{0.92}  \\
\textbf{CNN1d}  & 19.45 & 26.56 & 1.72 & 0 & 0 & 1.00  \\
\textbf{BiLSTM} & 1.35 & 4.69 & 1.93 & 5.45 & 18.75 & 0.96 \\
\bottomrule
\end{tabular}
\end{table}

\subsection{Proposed Defense Method}
Depending on the above results, it is evident to conclude that the robustness varies with the differences of the models, when targeted attacks are conducted. However, we go a step further to discover what plays a major role in this difference. By comparing AlexNet with other models such as WDCNN, we found that the proper use of data normalization can facilitate convergence, and defend against the potential attacks, because the robust models shown in Table \ref{tab:tab3} have all used data normalization. Based on the above conclusion, we combine their processing procedures and propose a defense method, as shown in Algorithm 1. The steps shown in Algorithm 1 can universally be used as a defensive layer for NN-based fault-diagnosis models.

\begin{figure}[t]
  \centering
  \includegraphics[width=\linewidth]{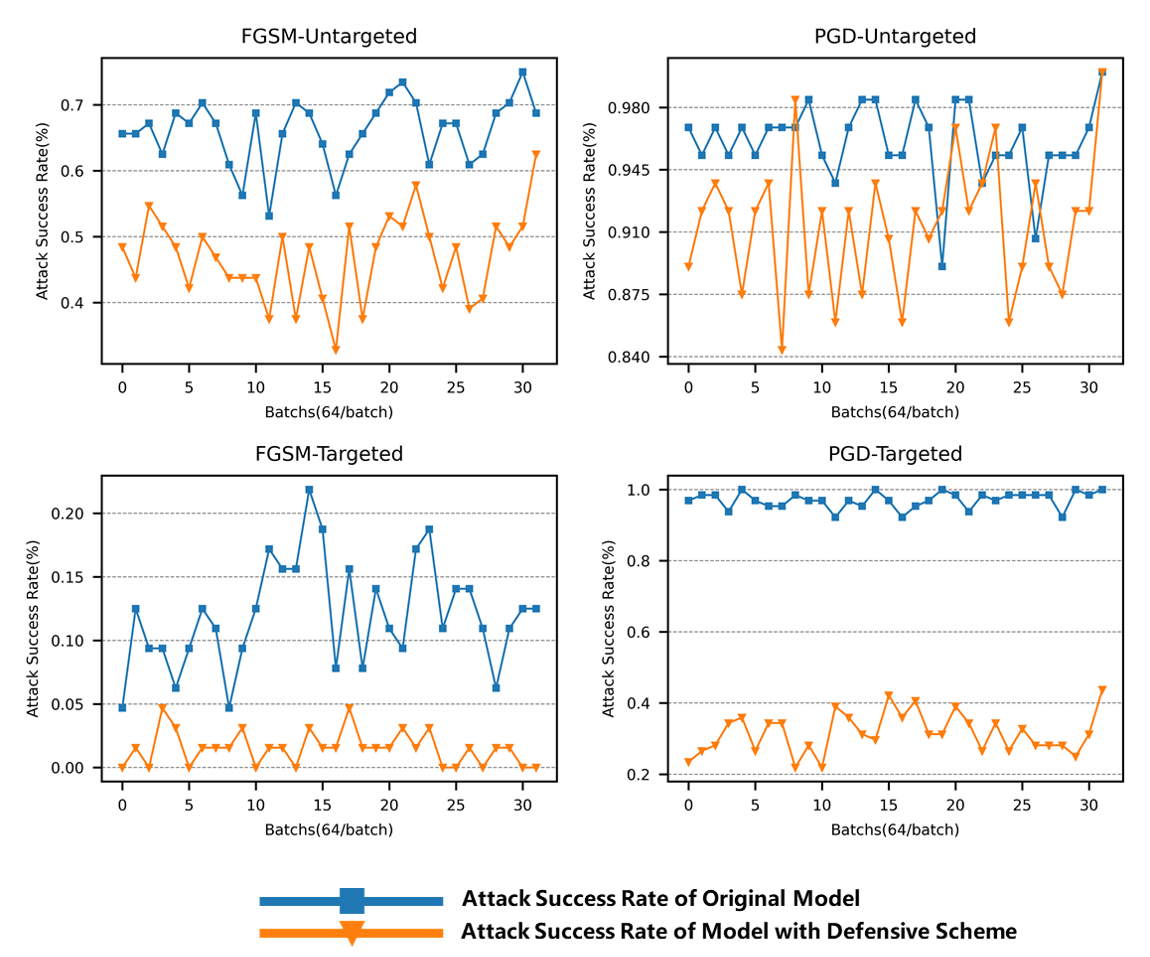}
  \caption{Comparison of the success rate between the original model and the one with defensive scheme.}
  \label{fig:pic5}
\end{figure}

To validate the defensive ability of the proposed method, we constructed this data process as a defensive layer, and added a defense layer to AlexNet before the first convolutional layer, to defend against the attacks of FGSM and PGD. The results are shown in Figure \ref{fig:pic5}: AlexNet with this defense strategy has more robustness, regardless of the attack methods. This demonstrates that our method can effectively defend against intentional attacks. Moreover, to avoid the influences of the model structure and hyperparameters, we remove the data normalization of other models to contrast them with the original models. We found that the success rates had varying degrees of increasements after being remove the data normalization.

\begin{algorithm}[h]
  \caption{Proposed defensive method DN} 
  \begin{algorithmic}[1]
    \Require
      Values of $x$ over a batch:$\beta=\{x_1 .. x_m\}$;
      Parameters to be learned:$\gamma,\delta$;
    \Ensure
      $\{ y= DN_{\gamma,\delta}(x);\}$
       \State set $sum_1\leftarrow 0,sum_2\leftarrow 0$;
        \For{$i = 1$; $i<m$; $i++$ }
            \State $sum_1\leftarrow sum_1+x_i$;
        \EndFor
      \State $\mu_{\beta}\leftarrow sum_1/m$;
      \For{$i = 1$; $i<m$; $i++$ }
            \State $sum_2\leftarrow sum_2+(x_i-\mu_{\beta})^2$;
      \EndFor
      \State $\sigma_{\beta}^2\leftarrow sum_2/m$;
      \For{$i = 1$; $i<m$; $i++$ }
            \State $\widehat{x_i}\leftarrow \frac{x_i-\mu_{\beta}}{\sqrt{\sigma_{\beta}^2-\epsilon}}$;
            \State $y_i\leftarrow \gamma \times \widehat{x_i} +\delta $;
      \EndFor
  \end{algorithmic}
\end{algorithm}

From the result of the proposed scheme, it is obvious that DN works well. In fact, the similar phenomenon was also observed recently in the image domain \cite{ref23}, but the defensive effect of similar method increases greatly in machinery vibration signals domain. To explain this result, we found that DN can map inputs to other fields, amplifying small sample differences due to data processing. Therefore, the amplified differences can act as crucial features to distinguish classes. The models without defensive layers, however, lost their robustness for the differences between the adversarial and the original inputs are small. Additionally, DN can also enhance the robustness of models by mitigating the instability during adversarial training \cite{ref24}. Future works can be done to extend the application of DN in other fields.

\section{Conclusion and future work}

We proposed an adversarial example of machinery vibration signals under untargeted and targeted conditions. The distortion measure was redefined for different restrictions of vibration signals. Our experimental results indicate that it’s possible to conduct the attacks using existing methods and achieve high attack success rates without being discovered. Additionally, the proper use of data normalization can not only help the models converge quickly but also defend against the potential adversarial attacks effectively. 

We will further investigate the black-box adversarial attack of machinery vibration signals and determine more effective measures to defend against various attacks for eliminating the potential risks of the trouble caused by these attacks. We hope that more researchers in related fields can pay attention to these problems, since the reliability of the fault diagnosis system is always an important issue when the system is applied in real scenarios.

\clearpage
\bibliographystyle{IEEEtran}

\end{document}